# Large Frequency Change with Thickness in Interlayer Breathing Mode - Significant Interlayer Interactions in Few Layer Black Phosphorus

Xin Luo,[⊥,&] Xin Lu,[†] Gavin Kok Wai Koon,[⊥] Antonio H. Castro Neto,[⊥,#] Barbaros Özyilmaz,[⊥,]* Qihua Xiong,[†,Ŧ,]* and Su Ying Quek[⊥,&,]*

[⊥]Department of Physics, Centre for Advanced 2D Materials and Graphene Research Centre, Faculty of Science, National University of Singapore, 2 Science Drive 3, Singapore 117551

[&]Institute of High Performance Computing, 1 Fusionopolis Way, #16-16 Connexis, Singapore 138632

[†]Division of Physics and Applied Physics, School of Physical and Mathematical Sciences, Nanyang Technological University, Singapore 637371

[Ŧ]NOVITAS, Nanoelectronics Centre of Excellence, School of Electrical and Electronic Engineering, Nanyang Technological University, Singapore 639798

[#]Boston University, 590 Commonwealth Ave., Boston MA 02215

*To whom correspondence should be addressed. Email: phyqsy@nus.edu.sg (theory) qihua@ntu.edu.sg (Raman spectroscopy experiments) barbaros@nus.edu.sg (sample preparation)

**ABSTRACT**

Bulk black phosphorus (BP) consists of puckered layers of phosphorus atoms. Few-layer BP, obtained from bulk BP by exfoliation, is an emerging candidate as a channel material in post-silicon electronics. A deep understanding of its physical properties and its full range of applications are still being uncovered. In this paper, we present a theoretical and experimental investigation of phonon properties in few-layer BP, focusing on the low-frequency regime corresponding to interlayer vibrational modes. We show that the interlayer breathing mode $A^3_g$ shows a large redshift with increasing thickness; the experimental and theoretical results agreeing well. This thickness dependence is two times larger than that in the chalcogenide materials such as few-layer $MoS_2$ and $WSe_2$, because of the significantly larger interlayer force constant and smaller atomic mass in BP. The derived interlayer out-of-plane force constant is about 50% larger than that in graphene and $MoS_2$. We show that this large interlayer force constant arises from the sizable covalent interaction between phosphorus atoms in adjacent layers, and that interlayer interactions are not merely of the weak van der Waals type. These significant interlayer interactions are consistent with the known surface reactivity of BP, and have been shown to be important for electric-field induced formation of Dirac cones in thin film BP.

**Main text:**

Two dimensional (2D) layered materials have received much attention in recent years due to their unique properties and potential applications in nanoelectronics, nano-optoelectronics, and in nanoelectromechanical systems.[1] Starting with graphene, the range of 2D materials has expanded to include 2D transition metal dichalcogenides, such as $MoS_2$, and most recently to the newly discovered 2D black phosphorus (BP). Single layer BP (phosphorene) has been investigated as a potential candidate in post-silicon electronics due to its semiconducting nature and high hole mobility.[2,3] However, the full extent of the properties and applications of 2D BP is still being uncovered. One common feature of all 2D materials is that single layers stack up to form layers in the bulk and in thin films. The interlayer interactions are commonly thought to be of the non-local weak van der Waals type. However, it has been found that interlayer interactions beyond the weak van der Waals type are important in $MoS_2$.[4] In particular the well-known direct to indirect band gap transition, as the thickness of $MoS_2$ increases from a monolayer, arises from Coulombic repulsion between sulphur atoms in adjacent layers.[4,5] Coulombic attraction between S and Mo atoms in adjacent layers of bulk $MoS_2$ are also responsible for the Davydov splitting in the ($E_{1u}^2$, $E_{2g}^1$) conjugate frequency modes,[6] while short-range interlayer interactions, possibly of covalent nature, result in enhanced surface force constants that give rise to anomalous frequency trends in 2D $MoS_2$[6] and $WSe_2$[7]. Unlike the transitional metal dichalcogenides, graphene and phosphorene are monotypic materials, so that no Coulombic interactions are expected.

However, in this work, we find that sizable covalent interactions exist between phosphorene layers – larger than that in graphene and $MoS_2$, resulting in significantly larger interlayer force constants and a large change in the interlayer breathing mode frequency as the thickness of 2D BP increases. Our results indicate that interlayer interactions in layered BP are not simply of the weak van der Waals type, and indicate the need for renewed effort to understand the implications of interlayer interactions on the properties of 2D BP.

Interlayer force constants in 2D layered materials can be theoretically computed, and experimentally determined from low frequency Raman spectra, as has previously been seen in multilayer graphene and dichalcogenide materials.[8, 9] We present first principles, density functional perturbation theory (DFPT) (see Methods for details) calculations for phonons and non-resonant Raman intensities in few-layer BP, focusing on the low frequency regime for interlayer modes. A linear chain model, which fits well to the DFPT results, is used to predict phonon frequencies for samples thicker than 7 layers. These results are compared with experimental data from Raman spectroscopy, which clearly show a low frequency interlayer breathing mode, $A^3_g$, that has thus far been out of the range of previous Raman experiments on few layer BP.

We begin our discussion with a description of the lattice and symmetries of bulk and few-layer BP, in order to understand the Raman selection rules of this system. Bulk BP, the most thermodynamically stable form of phosphorus at room temperature and pressure, is a narrow bandgap semiconductor with a direct bandgap of 0.3 eV.[10, 11] The bulk structure belongs to the orthorhombic lattice and consists of puckered layers of atoms as shown in Figure 1. A primitive unit cell contains four atoms in positions $\pm(0, \mu b, \nu c)$ and $\pm(a/2, (\mu+1/2)b, -\nu c)$.[12]

Table 1 summarizes the fully optimized lattice constants and internal coordinates obtained using different exchange-correlation functionals, together with the experimental data.[12] LDA stands for the local density approximation to the exchange-correlation functional, PBE is a gradient-corrected approximation[13], while VdW stands for the non-local VdW-DF2-c09 functional.[14, 15] In general, the interlayer distances in Van der Waals layered solids are quite well predicted using LDA (for fortuitous reasons) and VdW-DF2-c09 functionals[8, 9, 16] while PBE overestimates these distances, as can be seen also in Table 1. PBE-D2 adds the dispersion energy to the PBE functional with optimized dispersion coefficients in the empirical pair-wise force field expression.[17] By comparing the lattice parameters in phosphorene and bulk BP, we find that the lattice constant $a$ is smaller than bulk while the lattice constant $b$ is larger. Among the different functionals, the PBE-D2 and VdW functionals best describe the experimental bulk BP lattice parameters, while LDA also performs reasonably well.[12] On the other hand, as we shall see later, the interlayer phonon frequencies are best described using LDA. This is similar to the case for transition metal dichalcogenides,[8] and indicates that while the VdW functionals are well-optimized to give the structural parameters, they do not perform as well as the LDA for interlayer vibrational properties. Henceforth, we report the LDA results, unless otherwise stated.

The symmetry of the bulk BP crystal can be described by the *Cmca* space group

($D_{2h}^{18}$), and the irreducible representation of the phonon modes at the center of the Brillouin zone is $\Gamma_{bulk} = 2A_g + 2B_{3g} + 2B_{1u} + 2B_{2u} + A_u + B_{1g} + B_{2g} + B_{3u}$. Among the twelve vibration modes, there are six Raman-active modes ($B_{1g}$, $B_{2g}$, $2A_g$ and $2B_{3g}$), five infrared-active modes ($2B_{1u}$, $2B_{2u}$ and $B_{3u}$) and one optically inactive mode $A_u$. When the system goes from 3D to 2D, the translational symmetry along the z-axis is absent. As a result, thin films with an odd and even number of layers belong to the space groups *Pmna* ($D_{2h}^7$) and *Pbcm* ($D_{2h}^{11}$), respectively.[18] Consequently, the thin films' irreducible representation of phonon modes at the Gamma point is $\Gamma_{n-layers}$ =$n(2A_g + 2B_{3g} + 2B_{1u} + 2B_{2u} + A_u + B_{1g} + B_{2g} + B_{3u})$. Since the thin film and bulk BP have the same $D_{2h}$ point group with inversion symmetry, they share the same character table. According to group theory, the Raman tensors of the Raman active $A_g$, $B_{1g}$, $B_{2g}$ and $B_{3g}$ modes are:[18]

$$A_g : \begin{pmatrix} a & 0 & 0 \\ 0 & a & 0 \\ 0 & 0 & a \end{pmatrix}, B_{1g} : \begin{pmatrix} 0 & a & 0 \\ a & 0 & 0 \\ 0 & 0 & 0 \end{pmatrix}, B_{2g} : \begin{pmatrix} 0 & 0 & a \\ 0 & 0 & 0 \\ a & 0 & 0 \end{pmatrix}, B_{3g} : \begin{pmatrix} 0 & 0 & 0 \\ 0 & 0 & a \\ 0 & a & 0 \end{pmatrix}$$

Therefore, the $A_g$, $B_{1g}$, $B_{2g}$ and $B_{3g}$ modes can be detected using Raman spectroscopy under the $\bar{z}(xx)z$, $\bar{z}(xy)z$, $\bar{z}(xz)z$ and $\bar{z}(yz)z$ polarization configuration, respectively. The notation of $\bar{z}(xy)z$ means that the incident light with polarization parallel to the *x* axis propagates along the $\bar{z}$ axis and the scattered light polarized parallel to the *y* axis propagates along the $z$ axis.

Previous Raman studies of BP have focused on the high frequency intra-layer vibrations such as the $A_g^1$, $A_g^2$ and $B_{2g}$ modes,[2,19] while the interlayer vibration modes are still largely unknown. From our LDA calculations, we found that there are several branches of interlayer modes in the low frequency range. Their frequencies and Raman activities are summarized in Table 2. The interlayer vibrational modes in few-layer BP are derived from the acoustic mode of phosphorene, and there will be N-1 derived interlayer modes in N-layer BP. Among these interlayer modes, the Raman active breathing (out-of-plane) modes are described by the $A_g$ representation, which can be observed under the $\bar{z}(xx)z$ configuration. To distinguish these from the intra-layer $A_g$ mode, we shall use the notation $A_g^3$ to represent the largest Raman intensity interlayer breathing mode. The $A_g^3$ mode can be found in N layer BP (N≥2), with the lowest frequency in all derived breathing modes. Besides the $A_g^3$ mode, there are still other trends of breathing modes $A_g$, which will appear when the layer thickness is ≥ 4 layers, but with much smaller Raman intensity. Similar to the other layered materials, there are also interlayer shear modes in few layer BP, as shown in Table 2. In contrast to the degenerate shear modes in few-layer graphene and MoS$_2$,[8,9] the degeneracy is broken in few layer BP, resulting in two different sets of shear modes. The shear mode with atomic displacement along the armchair direction is represented as the $B_{3g}$ mode, which can be observed in the $\bar{z}(yz)z$ polarization configuration. The shear mode with atomic displacements along the zigzag edge is the $B_{3g}^3$ mode, which can be detected in $\bar{z}(xz)z$ configuration. We use the notation $B_{3g}^3$ and $B_{2g}^2$ to label the highest intensity interlayer shear modes in the armchair and zigzag directions respectively.

Figure 2 shows the atomic displacements of the $A_g^3$, $B_{3g}^3$, $B_{2g}^2$ and other Raman-active interlayer modes. Since the whole layer is moving as a unit, the arrow indicates the movement of that particular layer. Although the atomic displacements of the three interlayer modes are along different directions, they have similar patterns and all exhibit red shifts with increasing thickness. These red shifts can be understood from the atomic displacements (Figure 2) - the total relative change in atomic displacement decreases with increasing thickness, so that the total accumulated force and therefore the frequency also decreases. From the calculated Raman intensities listed in Table 2, we note that the Raman intensities of the $B_{3g}^3$ and $B_{2g}^2$ mode are much smaller than the $A_g^3$ mode, making it more difficult to detect these modes.

In our calculations, the Raman intensity of the $A_g^3$ mode is comparable in value to the high frequency intralayer $A_g^1$ and $A_g^2$ modes. Thus, we expect that the $A_g^3$ mode can also be observed in experiment. We therefore focus on this mode and show in Figure 3(a) the predicted ultra-low frequency Raman spectra of 1-7 layer and bulk BP under the $\bar{z}(xx)z$ configuration. As expected, this low frequency breathing mode $A_g^3$ is absent in the single layer due to its interlayer vibrational feature. On the other hand, the bulk limit of the $A_g^3$ mode is not Raman active due to the cancelation of accumulated bond polarizability under the periodic conditions. Interestingly, we find that the frequency of the $A_g^3$ mode changes significantly with thickness, much more than that in the previously reported $A_g^2$ mode for BP,[2, 19] making it possible to determine the BP film thickness using Raman spectroscopy for the $A_g^3$ mode.

Inspired by above theoretical predictions, we performed Raman scattering experiments to detect the ultra-low frequency vibrational modes. The sample thickness is determined using a combination of atomic force microscopy (AFM) and optical contrast imaging – the former is more reliable for the thicker samples and the latter for the thinner samples. An example of the thickness determination using AFM is shown in Figure 4(b-c). Although our theoretical thickness for single layer phosphorene is 0.52 nm, we use 0.6 nm for calibrating the AFM measurements, this calibration is consistent with previous studies.[2] Figure 4d shows an 11 layer BP sample, and the rough surface is due to possible adsorbates from environment.[20] For thinner samples, the rough surface makes it difficult to ascertain the sample thickness accurately. On the other hand, the band gap is very sensitive to thickness for thinner samples.[21] We therefore relied also on optical contrast images[22, 23] to determine the sample thickness more accurately. The Raman spectra are excited by the 633 nm laser in a back scattering configuration with the incident light polarized along $x$ axis. Experimental results are shown in Figure 4 (and Figure 3b). In the main text, we show only the spectra acquired with non-polarized light, as the signal-to-noise ratio is much poorer when polarization configurations are considered. The spectra for different polarization configurations were also acquired and are shown in Figure S1 for information. Samples down to 2-3 layers thin were prepared. However, we were only able to obtain Raman spectra for samples thicker than 3 layers. This is because even a laser power of 37 μW and 3.5 μW will cause damage to the 3 layer BP sample, as shown in the Figure S2 (typically, a laser power of 0.35mW is used to detect the interlayer modes), while in the 4 layer BP sample, a 90 μW laser had to be used.

There are three frequency trends below 100 cm$^{-1}$ in Figure 4(a) - the lowest frequency modes guided by the red dashed line is the $A_g^3$ breathing mode predicted in our calculation, which shows a large redshift with increasing thickness. The frequencies of this observed lowest frequency modes are in good agreement with our LDA predicted $A_g^3$ breathing mode, as shown in Figure 3(b), with direct comparison between experiment and LDA for 4-7 layer BP. The Raman-active breathing mode with the second lowest frequency (we call this $A_g^4$) corresponds well to the highest frequency peak highlighted for 5L BP (LDA: 83.7cm$^{-1}$; experimental peak position: 85cm$^{-1}$), but according to our calculations, this mode red shifts as the thickness increases, to 65.4cm$^{-1}$ for 7L (LDA), and 37.5cm$^{-1}$ for 13L (linear chain model), in contrast to the measured trends which blue shift with increasing thickness in the 70-90 cm$^{-1}$ range for 5-13 layers. Interestingly, however, a peak at 38cm$^{-1}$ is experimentally observed in 13 layer BP (corresponding well to $A_g^4$), as can be seen in Figure 4e. Therefore, the $A_g^4$ peak seems to be observed for 5 and 13 layer BP, while the modes with highlighted trends in the 70 to 90 cm$^{-1}$ range are still largely unknown.

With decreasing thickness, the full width at half maximum of the $A_g^3$ mode becomes larger. This trend is commonly observed in other materials, and can be explained by increased boundary and impurity scattering in thinner samples,[24] as well as the commonly accepted RWL phonon confinement theory.[25] We note that in several (but not all) of our spectra, there is an asymmetric broadening of the $A_g^3$ breathing mode. By performing multi-Lorentzian fitting on the asymmetric peaks (Figure 4e), we find that the asymmetry is due to an extra peak X1, which shows a red shift with increasing thickness, changing from 20.2 cm$^{-1}$ to 16.4 cm$^{-1}$ when the thickness increases from 6 to 8 layers. The X1 peak may come from the second lowest frequency interlayer $B_{3g}$ shear mode (we call this $B_{3g}^4$), which is 2-4 cm$^{-1}$ lower than the $A_g^3$ mode from our LDA calculations.

We now focus again on the lowest frequency $A_g^3$ mode, which has the highest predicted Raman intensity among the low frequency interlayer modes, and also shows good correspondence between experiment and theory. We analyze our results by using a free standing linear chain model based on nearest-neighbor interactions to fit the LDA results. In the linear chain model, we consider the puckered layer as one single unit, and it is connected with neighboring layers (units) by force constants. The linear chain model has been used successfully to describe the experimental and DFT interlayer phonon modes in other layered materials.[8,9] The phonon frequency of the interlayer mode can be obtained by diagonalizing the dynamical matrix of the finite N layer BP. For the above mentioned $A_g^3$ breathing mode, its thickness dependent frequency can be simply described by:

$$\omega_\alpha = \omega_0 \sqrt{2} \sin\left(\frac{\pi}{2N}\right) \quad (1)$$

$\omega_o$ corresponds to the frequency of the interlayer breathing mode in 2 layer BP, that is 75.0 cm$^{-1}$ in our calculation. The curve of equation (1) shown as the red line in Figure 3(b) fits both the LDA and experimentally measured frequency well. The good fitting suggests that the interlayer interactions are dominated by the interactions between nearest-neighboring layers. Similar to few layer graphene and MoS$_2$,[8,9] adsorbates or substrate effects seem to have negligible influence on the frequencies of the interlayer breathing mode. In Fig. 3(b), we show also the VdW-calculated frequencies, which do not follow the linear chain model well, in contrast to the experimental and LDA results. Using the above expression with the $\omega_o$ obtained from DFT LDA, the thickness of the thin film BP can be rapidly determined from the low frequency breathing mode. Furthermore, from the linear-chain model, $\omega_o$ can be expressed as $\omega_0 = \frac{1}{\sqrt{2}\pi c}\sqrt{\frac{k_z}{\mu}}$, where $\mu$ is the mass per unit area of the few layer BP, $c$ is the speed of light in cm/s, and $k_z$ is the out of plane interlayer force constants. The thus derived out-of-plane force constant $k_z$ (14.1×10$^{19}$ N/m$^3$) is related with the elastic modulus of few-layer BP by $C_{33}=k_z l$ with $l$ being the distance between the centers of mass for adjacent puckered layers. As a result, the calculated elastic modulus of $C_{33}$ is around 73.9 GPa, consistent with calculated values for the bulk.[26] The interlayer out-of-plane force constant $k_z$ is significantly larger than those derived in the same way for other materials ($k_z$ is 6.11×10$^{19}$, 9.6×10$^{19}$ and 9.3×10$^{19}$ N/m$^3$ for Bi$_2$Se$_3$,[27] graphite[28] and MoS$_2$,[8] respectively). Similarly, if we fit the linear chain model to our experimental data, we obtain $k_z$ of 12.3×10$^{19}$ N/m$^3$, larger than $k_z$ obtained by fitting the experimental frequencies of other materials.[8,27,29] The larger force constants contribute to a larger $\omega_o$ factor in equation (1), and thus a larger thickness dependence for the $A^3_g$ frequency. Specifically, the $A^3_g$ frequency changes 54 cm$^{-1}$ from 2-layer to 8-layer BP, two times more than that in the transition metal dichalcogenides such as few layer MoS$_2$ (31 cm$^{-1}$ from 2-layer to 8-layer) and WSe$_2$ (20 cm$^{-1}$ from 2-layer to 8-layer).[8] The above analysis can also be applied to the shear modes $B^3_{3g}$ and $B^2_{2g}$, and we can get a shear elastic modulus of 8.9 GPa and 20.4 GPa along the armchair and zigzag directions, respectively. The difference in the shear elastic moduli along the two in-plane directions reflects the strong anisotropic elastic properties of BP.

The larger force constant in BP suggests that the interlayer bonding may be different from that in the other well-studied layered materials like $MoS_2$ and Graphite. Previously, it has already been shown that Coulombic interaction can affect the phonon frequencies in thin film $MoS_2$,[6, 30, 31] while covalent interlayer interactions are also likely to explain the larger surface force constants, and resulting anomalous frequency trends in the same material.[6] BP is a monotypic material, where all phosphorus atoms are in principle in the same local environment (not considering surface effects). We therefore do not expect a significant Columbic interlayer interaction in BP. In order to ascertain the strength and nature of the interlayer covalent interactions, we use the VdW functional to compute the difference between the charge density of bilayer BP and the sum of charge densities of the component phosphorene layers (Figure. 5 a-c). As shown in Figure 5(c), we find a clear covalent character in the interlayer interaction between adjacent layers, *i.e.* the electrons are shared in the bonds between interlayers, with the largest electron density in the region between the nearest neighboring atoms in adjacent layers. Interestingly, while there is a clear directional character as described, the accumulated electron charge density is also slightly delocalized along the zigzag direction of the phosphorene lattice, giving rise to the strong anisotropic in-plane elastic constants. This interlayer covalent interaction is significantly larger than that in $MoS_2$ and graphene, as shown in Figure 6. Furthermore, compared with bilayer $MoS_2$ and graphene, the binding energy of bilayer BP is 2 times larger in both LDA and VdW calculations (Figure 5d). Since the interlayer force is proportional to the curvature of the binding energy versus distance, we also plot the derivative of the binding energy as a function of the separation distance in Figure 5(e), which shows a larger curvature in BP compared with bilayer MoS$_2$ and graphene.

The large interlayer covalent character is on hindsight not surprising, given that the phosphorene atoms each have an electron lone pair sticking into the interlayer vacuum region (phosphorus has five valence electrons, only three of which are involved in forming covalent bonds). Furthermore, this sizable interlayer covalent interaction is consistent with the known surface reactivity of BP[32, 33] as well as the large interlayer band dispersion and small excitonic effects in bulk BP[34, 35] On the other hand, it has also recently been predicted that electric fields applied perpendicularly to the BP thin films can result in a topological phase transition involving the formation of Dirac cones in the band structure.[36, 37] We have explained this phenomenon using a tight-binding model, and have shown that the interlayer interactions are a necessary condition for the formation of the Dirac cones.[37] Similarly, we suggest that the formation of topological phases in bulk BP under pressure[38] may also be related to the large covalent interlayer interactions – a topic worth further investigating.

In conclusion, we have theoretically predicted and experimentally measured the Raman spectra of interlayer modes in few layer BP to gain insights into the interlayer force constants and interlayer interactions in BP. We demonstrate that the interlayer breathing modes have a large red shift in frequency with increasing thickness, corresponding to an unusually large interlayer force constant. We show that the interlayer force constants arise from sizable interlayer covalent interactions in BP. We have discussed a few known implications of this large interlayer interaction, and suggest that this unusual interlayer interaction presents a playground for further new discoveries.

**Methods:**

**First Principles Calculations**

First principles calculations of vibrational Raman spectra are performed within density-functional theory (DFT) as implemented in the plane-wave pseudopotential code QUANTUM-ESPRESSO[39]. The local density approximation (LDA)[40] to the exchange-correlation functional is employed in the norm-conserving (NC)[41] pseudopotential throughout the Raman spectra calculation. For the purpose of comparison, a van der Waals functional (VdW-DF2),[15] with the Cooper's gradient correction on the exchange[14] and PBE calculation with dispersion correction in Grimme's scheme (PBE-D2) are also performed to calculate the lattice parameters. To get converged results, plane-wave kinetic energy cutoffs of 75 Ry are used for the wave functions. The slabs are separated by 16 Å of vacuum to prevent interactions between slabs (this value has been tested for convergence of phonon frequencies). A Monkhorst-Pack k-point mesh of 19×15×1 and 11×11×13 are used to sample the Brillouin Zones for the thin films and bulk systems, respectively. In the self-consistent calculation, the convergence threshold for energy is set to $10^{-9}$ eV. All the atomic coordinates and lattice constants are optimized with the Broyden–Fletcher–Goldfarb–Shanno (BFGS) quasi-Newton algorithm. The structures are considered as relaxed when the maximum component of the Hellmann-Feynman force acting on each ion is less than 0.003 eV/Å. With the optimized structures and self-consistent wave functions, the phonon spectra and Raman intensities are calculated within density-functional perturbation theory (DFPT) as introduced by Lazzeri and Mauri.[42] For the DFPT self-consistent iteration, we used a mixing factor of 0.1 and a high convergence threshold of $10^{-18}$ eV. In our DFPT calculations, we compute the static limit of the dielectric response self-consistently (including local field effects).[43] We ignore the frequency dependence of the dielectric matrix as well as the ionic contribution to the dielectric response. These approximations have been found to work very well in practice and are particularly justified in our case given the very small phonon frequencies that are of interest here.

**Synthesis**

The samples of few-layer black phosphorus are mechanically exfoliated from bulk crystals (purchased from Smart Elements) onto a Si substrate covered with thermally grown 300 nm of $SiO_2$. Prior to exfoliation, the substrate surface is cleaned and activated with $O_2$ plasma to enhance the yield. An optical microscope is utilized to locate the position of few-layer black phosphorus. The thickness of as-prepared samples is determined by both optical contrast and AFM measurements.

**Raman Spectroscopy**

Raman scattering spectroscopy measurements are carried out at room temperature using a micro-Raman spectrometer (Horiba JY-T64000) in a backscattering configuration. A Helium-Neon laser ($\lambda$=633 nm) has been used to excite the samples. The backscattered signal was collected through a 100× objective and dispersed by a 1800 g/mm grating under a triple subtractive mode, this setting can achieve a spectra resolution of ~1 cm$^{-1}$. To avoid the laser heating effect on the sample, laser power at the sample surface was less than 0.4 mW.

**ACKNOWLEDGEMENT**

S.Y.Q and X.L gratefully acknowledge the Singapore National Research Foundation (NRF) for funding under the NRF Fellowship (NRF-NRFF2013-07). The computations were performed on the cluster of NUS Graphene Research Centre and the A*STAR Computational Resource Center. Q.X. gratefully thanks Singapore National Research Foundation under the Fellowship (NRF-RF2009-06) and the Investigatorship (NRF-NRFI2015-03) and Ministry of Education via a tier2 grant (MOE2012-T2-2-086).

*Supporting Information Available:* Polarisation dependent measurement of few layer BP and Raman measurement on 3 layer BP samples. This material is available free of charge *via* the Internet at http://pubs.acs.org.

Table 1. Lattice parameters of bulk BP and phosphorene, the atomic positions in the orthorhombic lattice are specified by three lattice constants, *a, b, c*, and two structural parameters, $\mu$ and $\nu$. The equilibrium inter-layer distance *d* shown in Figure 1 is also given.

|  |  | *a* (Å) | *b* (Å) | *c* (Å) | $\mu$ | $\nu$ | *d* |
|---|---|---|---|---|---|---|---|
| LDA | bulk | 3.306 | 4.137 | 10.184 | 0.0731 | 0.1060 | 2.934 |
|  | 1L | 3.268 | 4.363 | - | 0.0813 | - | - |
| PBE | bulk | 3.305 | 4.563 | 11.312 | 0.0871 | 0.0935 | 3.540 |
|  | 1L | 3.298 | 4.624 | - | 0.0894 |  |  |
| VdW | bulk | 3.326 | 4.241 | 10.320 | 0.0743 | 0.1050 | 2.992 |
|  | 1L | 3.292 | 4.469 | - | 0.0836 | - | - |
| PBE-D2 | bulk | 3.322 | 4.430 | 10.475 | 0.0829 | 0.1019 | 3.101 |
|  | 1L | 3.313 | 4.507 | - | 0.0859 | - | - |
| Experiment | bulk | 3.313 | 4.374 | 10.473 | 0.0806 | 0.1034 | 3.071 |

Experiment is from Reference 12.

**Table 2. LDA calculated Γ point phonon frequencies (cm$^{-1}$) and relative Raman intensities (in parentheses) of the phonon modes in 1-7 layer BP. The intensity is normalized by the biggest value in each layer. The irreducible representation and Raman [R]/infrared [I] activity are also indicated.**

|  | Irr. | 1layer | 2layer | 3layer | 4layer | 5layer | 6layer | 7layer |
|---|---|---|---|---|---|---|---|---|
| Out of plane vibrations | $B_{1u}$ [I] | 0.0 (0) | 0.0 (0) | 0.0 (0) | 0.0 (0) | 0.0 (0) | 0.0 (0) | 0.0 (0) |
|  | $A_g$ [R] ($A^3_g$) |  | 75.0 (1.0) | 53.4 (1.0) | 40.7 (1.0) | 32.8 (1.0) | 27.6 (1.0) | 23.8 (1.0) |
|  | $B_{1u}$ [I] |  |  | 89.8 (0) | 74.1 (0) | 61.8 (0) | 52.9 (0) | 46 (0) |
|  | $A_g$ [R] ($A^4_g$) |  |  |  | 94.3 (0.01) | 83.7 (0.01) | 73.6 (0.03) | 65.4 (0.03) |
|  | $B_{1u}$ [I] |  |  |  |  | 96.5 (0) | 89.0 (0) | 80.9 (0) |
|  | $A_g$ [R] |  |  |  |  |  | 97.3 (0.003) | 92.2 (0.002) |
|  | $B_{1u}$ [I] |  |  |  |  |  |  | 98.0 (0) |
| In-plane (armchair direction) vibrations | $B_{2u}$ [I] | 0.0 (0) | 0.0 (0) | 0.0 (0) | 0.0 (0) | 0.0 (0) | 0.0 (0) | 0.0 (0) |
|  | $B_{3g}$ [R] |  | 26.0 (0.002) | 18.6 (0.01) | 13.9 (0.01) | 10.9 (0.02) | 9.5 (0.02) | 8.2 (0.02) |
|  | $B_{2u}$ [I] |  |  | 33.08 (0) | 24.2 (0) | 19.4 (0) | 18.6 (0) | 15.4 (0) |
|  | $B_{3g}$ [R] |  |  |  | 34.1 (0.001) | 26.7 (0.002) | 24.9 (0.002) | 21.6 (0.002) |
|  | $B_{2u}$ [I] |  |  |  |  | 33.8 (0) | 32.1 (0) | 26.5 (0) |
|  | $B_{3g}$ [R] |  |  |  |  |  | 35.0 (0.000) | 32.3 (0.000) |
|  | $B_{2u}$ [I] |  |  |  |  |  |  | 34.3 (0) |
| In-plane (zigzag direction) vibrations | $B_{3u}$ [I] | 0.0 (0) | 0.0 (0) | 0.0 (0) | 0.0 (0) | 0.0 (0) | 0.0 (0) | 0.0 (0) |
|  | $B_{2g}$ [R] |  | 39.4 (0.001) | 27.2 (0.001) | 20.6 (0.001) | 16.6 (0.001) | 13.7 (0.001) | 11.8 (0.001) |
|  | $B_{3u}$ [I] |  |  | 45.7 (0) | 38.1 (0) | 31.8 (0) | 26.6 (0) | 23.1 (0) |
|  | $B_{2g}$ [R] |  |  |  | 47.8 (0.000) | 43.0 (0.000) | 37.2 (0.000) | 33.1 (0.000) |
|  | $B_{3u}$ [I] |  |  |  |  | 49.2 (0) | 44.8 (0) | 41.0 (0) |
|  | $B_{2g}$ [R] |  |  |  |  |  | 49.0 (0.000) | 46.4 (0.000) |
|  | $B_{3u}$ [I] |  |  |  |  |  |  | 49.5 (0) |

**Figures:**

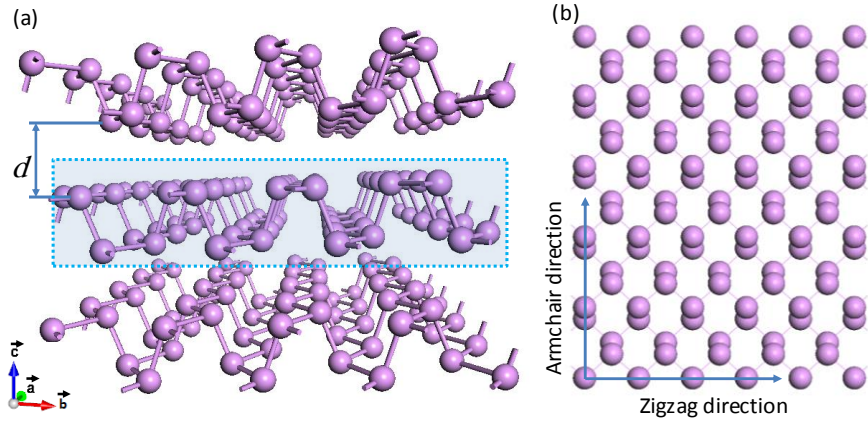

**Figure 1. (a) Structural illustration of black phosphorus and the equilibrium interlayer distance *d*. Phosphorene is a single-layer of the puckered sheet as depicted by the blue rectangle. (b) Top view of Phosphorene, with the zigzag and armchair directions indicated by blue arrows.**

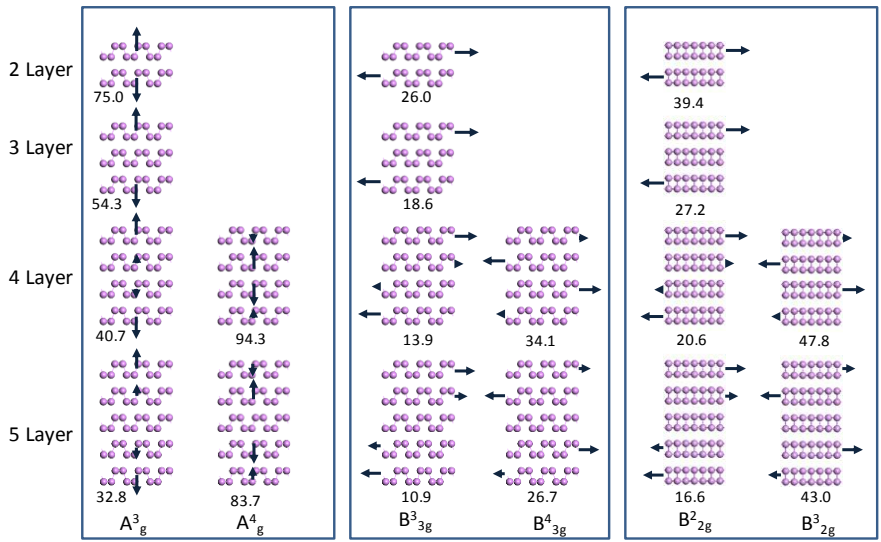

**Figure 2. The atomic displacements for the Raman active interlayer vibration modes in 2-5 layer BP; the frequency labelled below the displacement pattern is in the unit of cm$^{-1}$.**

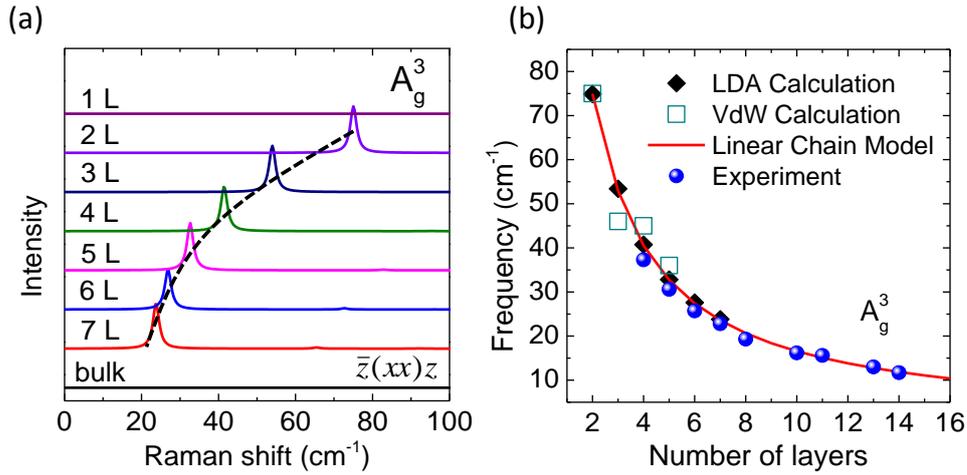

**Figure 3.** (a) DFT-LDA calculated Raman spectra of few layer BP in the ultra-low frequency range, with the black dashed lines guiding the Raman-active interlayer breathing mode. Note that the peaks are broadened artificially by Lorentzians. (b) Frequency evolution of DFT calculated and measured interlayer breathing mode as a function of the number of layers.

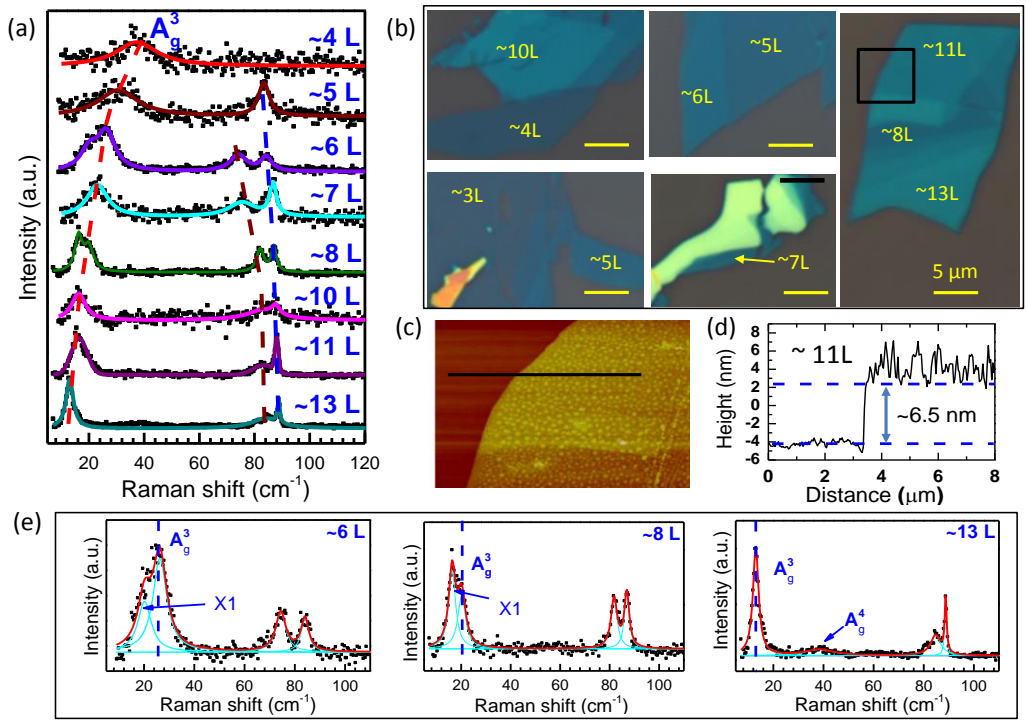

**Figure 4.** (a) Measured Raman spectra of few layer BP in the frequency range of 10-120 cm$^{-1}$. (b) Optical images of few layer BP samples on Si substrate capped with 300 nm SiO$_2$. Scale bars correspond to 5 μm. (c) AFM image of the **highlighted region (inside the black rectangle of Figure 4b) and (d) height profile as indicated by the black line in Figure 4c; the noise signal suggests surface absorption. (e) Zoomed-in Raman spectra of 6 layer, 8 layer and 13 layer BP, where the experimental data are shown in black dots, and red and blue lines are fitted results from multi-Lorentzian fitting.**

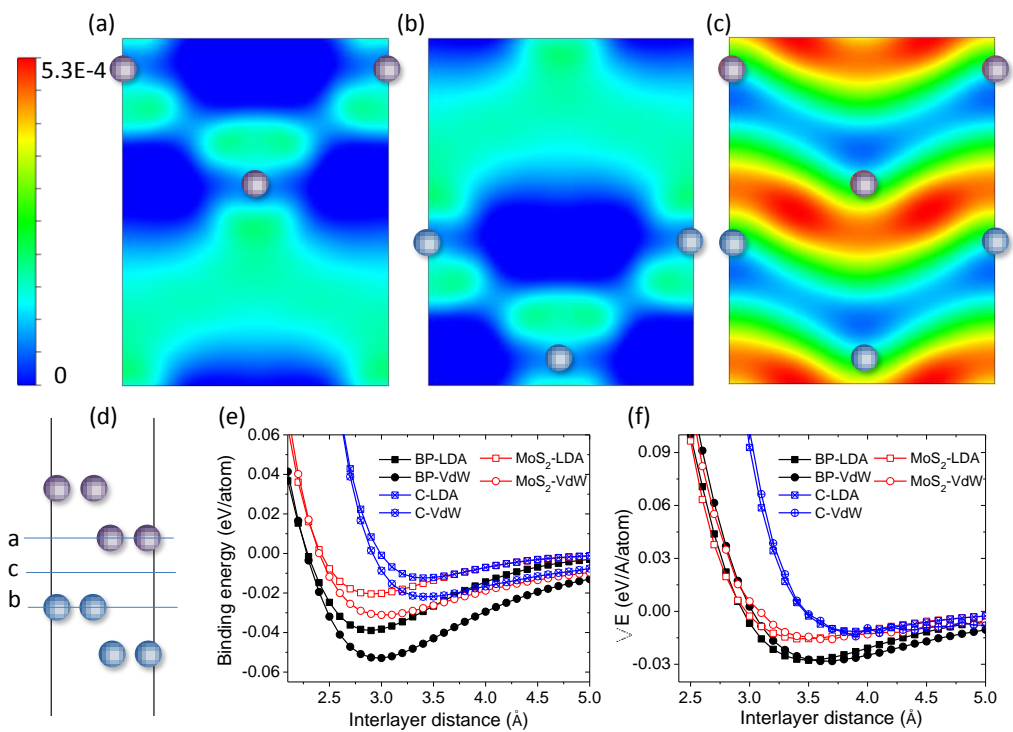

**Figure 5. VdW calculated 2D contour plots of the charge density difference between bilayer BP and the sum of the phosphorene components, the charge density is in $a_0^{-3}$ units. The plots are made for the plane cutting through the (a) second, (b) third atomic plane and (c) the middle of the interlayer gap, as illustrated in (d). The atoms nearest to the middle of the interlayer gap are projected on the cutting plane in (c), with grey and blue balls denoting the atoms in the second and third plane, respectively. (e) and (f) are the binding energy and second derivative of binding energy as a function of the interlayer separation distance.**

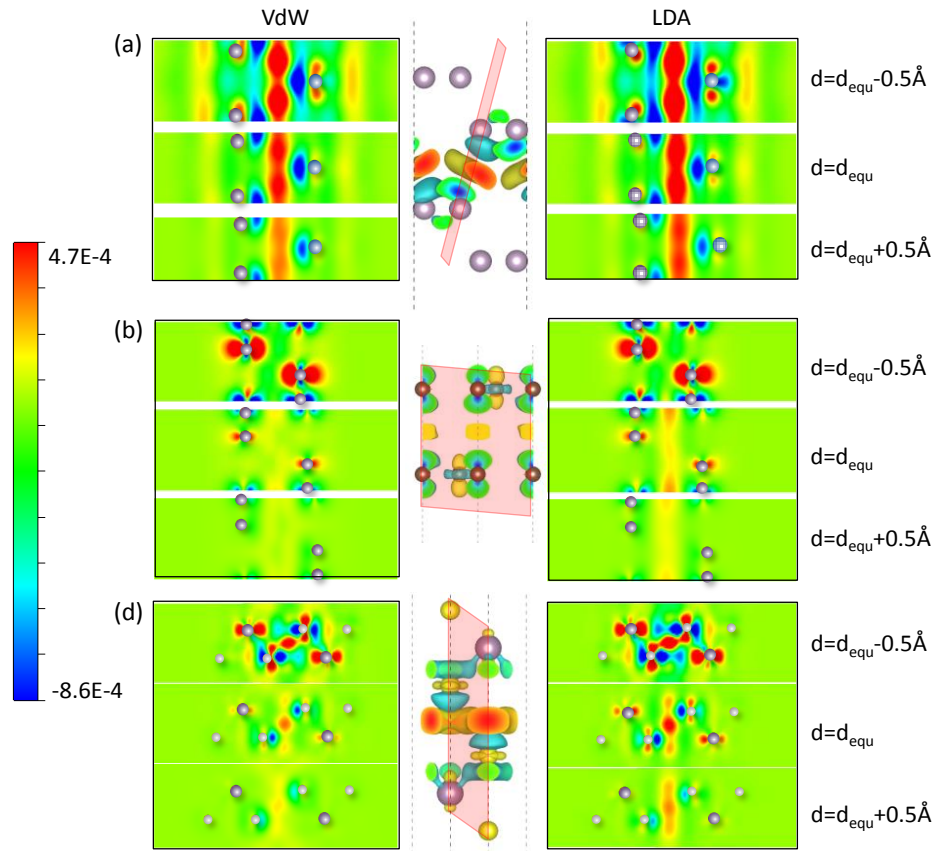

**Figure 6.** VdW and LDA calculated 2D charge density difference profiles for bilayer (a) BP, (b) graphene and (c) MoS$_2$ at different interlayer separation distances, the charge density is in $a_0^{-3}$ units. The contour plot in the middle of each system is obtained at the equilibrium interlayer distance. The cutting plane is shown in the transparent red color in the structure. The strong charge accumulation in BP indicates that the covalent interlayer interaction is significantly larger than that in MoS$_2$ and graphene.

TOC figure:

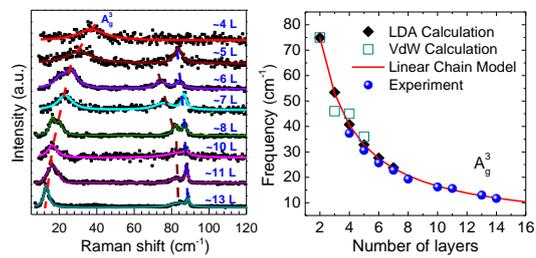